\documentstyle[epsf]{myl-aaii}

\newcommand{\rb}[1]{\raisebox{1.5ex}[-1.5ex]{#1}}
\newcommand{\reference}{Astron. Astrophys. 320, L5-L8 (1997)}

\begin{document}
   \thesaurus{13 % Sources as a function of wavelength
              (11.02.2; % BL Lacertae Objects: individual
               13.07.2) % Gamma rays: observations
              }

\title{Detection of $\gamma$-rays above 1.5 TeV from Mkn 501}
\author{
S.M. Bradbury \inst{1}
\and T. Deckers \inst{2}
\and  D. Petry \inst{1}
\and  A. Konopelko \inst{3}
\and  F. Aharonian \inst{3}
\and  A.G. Akhperjanian \inst{4}
\and J.A. Barrio \inst{1}\fnmsep\inst{5}
\and A.S. Beglarian \inst{4}
\and J.J.G. Beteta \inst{5}
\and J.L. Contreras \inst{5}
\and J. Cortina \inst{5}
\and A. Daum \inst{3}
\and E. Feigl \inst{1}
\and  J. Fernandez \inst{1}\fnmsep\inst{5}
\and V. Fonseca \inst{5}
\and A. Fra\ss \inst{3}
\and B. Funk \inst{6}
\and J.C. Gonzalez \inst{5}
\and V. Haustein \inst{7}
\and G. Heinzelmann \inst{7}
\and M. Hemberger \inst{3}
\and G. Hermann \inst{3}
\and M. Hess \inst{3}
\and A. Heusler \inst{3}
\and I. Holl \inst{1}
\and W. Hofmann \inst{3}
\and D. Horns \inst{7}
\and R. Kankanian \inst{3}\fnmsep\inst{4}
\and O. Kirstein \inst{2}
\and C. K\"ohler \inst{3}
\and D. Kranich  \inst{1}
\and H. Krawczynski \inst{7}
\and H. Kornmayer \inst{1}
\and H. Lampeitl \inst{3}
\and A. Lindner \inst{7}
\and E. Lorenz \inst{1}
\and N. Magnussen \inst{6}
\and H. Meyer \inst{6}
\and R. Mirzoyan \inst{1}\fnmsep\inst{5}\fnmsep\inst{4}
\and H. M\"oller \inst{6}
\and A. Moralejo \inst{5}
\and L. Padilla \inst{5}
\and M. Panter \inst{3}
\and R. Plaga \inst{1}
\and J. Prahl \inst{7}
\and C. Prosch \inst{1}
\and G. P\"uhlhofer \inst{3}
\and G. Rauterberg \inst{2}
\and W. Rhode \inst{6}
\and V. Sahakian \inst{4}
\and M. Samorski \inst{2}
\and J.A. Sanchez \inst{5}
\and D. Schmele \inst{7}
\and W. Stamm \inst{2}
\and M. Ulrich \inst{3}
\and H.J. V\"olk \inst{3}
\and S. Westerhoff \inst{6}
\and B. Wiebel-Sooth \inst{6}
\and C.A. Wiedner \inst{3}
\and M. Willmer \inst{2}
\and H. Wirth \inst{3}\newline
(HEGRA Collaboration)
}

\institute{Max-Planck-Institut f\"ur Physik, F\"ohringer Ring 6,
        D-80805 M\"unchen, Germany
\and Universit\"at Kiel, Inst. f\"ur Kernphysik,
       Olshausenstr.40, D-24118 Kiel, Germany
\and Max-Planck-Institut f\"ur Kernphysik, P.O. Box 103980,
        D-69029 Heidelberg, Germany
\and Yerevan Physics Institute, Yerevan, Armenia
\and Facultad de Ciencias Fisicas, Universidad Complutense,
         E-28040 Madrid, Spain
\and BUGH Wuppertal, Fachbereich Physik, Gau\ss str.20,
        D-42119 Wuppertal, Germany
\and Universit\"at Hamburg, II. Inst. f\"ur Experimentalphysik,
       Luruper Chaussee 149, D-22761 Hamburg, Germany
}
   \offprints{petry@hegra1.mppmu.mpg.de}

   \date{Received 29 November 1996; accepted 14 January 1997}

   \maketitle

\begin{abstract}
A detection of TeV $\gamma$-rays from Mkn 501 is reported,
based on observations made between March and August 1996 with the first HEGRA Cherenkov 
telescope (CT1). 
From the image analysis, 351 excess candidate $\gamma$-ray events are
obtained from the 147 h dataset. The statistical significance of the excess is 5.2 $\sigma$.
The average excess rate is $2.4 \pm 0.5$ h$^{-1}$ above the $\approx$ 1.5 TeV
threshold of CT1.
Under the assumption that the spectrum of Mkn 501 follows a power law
we find a differential spectral index
of 2.6$\pm0.5$ and obtain a time-averaged integral flux above 1.5 TeV of 
$2.3 (\pm0.4)_{\mathrm{Stat}} (+1.5-0.6)_{\mathrm{Syst}} \times 10^{-12}$ cm$^{-2}$s$^{-1}$.
Comparison with our near contemporary observations of the Crab Nebula, used as
a standard candle to test CT1 after upgrading to a 127 pixel camera, indicates that 
Mkn 501 has a spectrum similar to that of the Crab Nebula above 1.5 TeV.
The integral flux above 1.5 TeV from Mkn 501 is found to have been between 2.2 and 3.6 times 
smaller than that from the Crab Nebula. HEGRA is the second experiment to have detected Mkn 501 in the TeV range. 

\vspace{-0.3cm}
\vspace{-0.6cm}
\end{abstract}
\vspace{-0.5cm}
\section{ Introduction }
\vspace{-0.1cm}

Mkn 501, a relatively close BL Lac object (redshift $z$ = 0.034), became a target
for observation in the VHE waveband as a result of the unexpected emergence of AGN
as a significant sub-class of the high energy $\gamma$-ray sources identified by
EGRET onboard the Compton Gamma Ray Observatory (von Montigny et al. \cite{montigny}). 
Using the atmospheric Cherenkov technique, the Whipple group observed
$\gamma$ radiation of between 0.5 and 1.5 TeV from the BL Lac object Mkn 421
(Punch et al. \cite{punch})
and a confirmatory detection by two independent telescopes of the HEGRA Collaboration 
helped to establish its status as a VHE source (Petry et al. \cite{421paper} = {\it Paper I}).
The Whipple group has since surveyed more than 30 AGN with similar characteristics
(Kerrick et al. \cite{kerrick-b}).
Of these, one, Mkn 501, has been reported to emit above 300 GeV, at an average flux level
of $8 \times 10^{-12}$ cm$^{-2}$s$^{-1}$ measured between March and July 1995.
This was approximately 20$\%$ of the average Mkn 421 flux (Quinn et al. \cite{quinn}). Mkn 421 has  
been detected by EGRET whereas Mkn 501 has not.
\vspace{-0.08cm}

The HEGRA Collaboration's imaging air Cherenkov telescopes are part of its
cosmic ray detector complex (e.g. Rhode et al. \cite{rhode})
at the Observatorio del Roque de los
Muchachos on the Canary Island of La Palma (28.75$^\circ$ N, 17.89$^\circ$ W,
2200 m a.s.l.). The first telescope, CT1, is described in detail in
Mirzoyan et al. (\cite{mirzoyan-a}) and Rauterberg et al. (\cite{rauterberg}).
Its energy threshold is approx. 1.5 TeV. It has a 5\,m$^2$ reflector as opposed to the 
8.5\,m$^2$ dishes of the 5 telescope system (CT2-6) now nearing completion (Hermann \cite{hermann}).
CT2 and CT3 were targeting other objects whilst CT1 was chosen to monitor Mkn 501.

\section{Observations and data analysis}
\label{analysis}

Between March and August 1996 Mkn 501 was observed in tracking mode (rather than taking interleaved ON- and OFF-source
runs) to obtain maximum exposure time. 147 hours of good quality data were obtained  
at zenith angles $\theta$ between 11$^\circ$ and 25$^\circ$. 
Atmospheric extinction measurements from
the Carlsberg Automatic Meridian Circle near the HEGRA site were used as a guide to data quality.
Analysis proceeded in three steps: (1) flat-fielding and calibration,
(2) filtering to obtain a dataset of showers with well determined image parameters
(for a definition see e.g. Reynolds et al. \cite{reynolds}),
(3) selection of $\gamma$ candidates using cuts on the image parameters. For a more detailed description of 
our image analysis methods see Paper I and references therein.

\begin{enumerate}

\item{    
      Calibration and flat fielding are based on regular measurements
      of the pedestals and the relative photomultiplier gains.}
\item{CT1 carries a 127 photomultiplier camera
faced with hollow hexagonal light guides of 0.25$^\circ$ diameter. In addition to a 
hardware trigger condition of any $\ge$ 2 out of 127 pixels fired, a software trigger 
condition of any $\ge$ 2 out of 91 pixels above 16 photoelectrons was applied to the calibrated signals to exclude
camera-edge events with incomplete images.
Events recorded
under poor telescope positioning were rejected leaving a mean absolute pointing error of $<$0.1$^\circ$.

}
\item{
A series of image parameter cuts was applied which reject events of probable hadronic origin leaving a sample
of $\gamma$-shower candidates. For our detection of Mkn 421 (Paper I), neither 
observations of the Crab Nebula nor all parameters for the
Monte Carlo optimization of the cuts were available for the new camera, therefore we used the set of cuts
developed for a 91 pixel camera with similar resolution as described in Reynolds et al. (\cite{reynolds}): 
\vspace{-0.2cm}
\begin{center}
\begin{tabular}{c}
        $ 0.51^\circ < \mathrm{DIST} < 1.1^\circ $\\ 
   $ 0.07^\circ < \mathrm{WIDTH} < 0.15^\circ $ \\
   $ 0.16^\circ < \mathrm{LENGTH} < 0.30^\circ $\\ 
   $ \mathrm{ALPHA} < 10^\circ $ \\
\end{tabular}
\end{center}

In addition, a cut CONC $ > 0.4 $ was applied. We continued to use these previously successful cuts for our
analysis of Mkn 501. Monte Carlo data now being available, we can calculate
the flux from Mkn 501 by comparison with this simulated data which has not also been
used in optimisation of the cuts (see Section \ref{mcstudy}).
}
\end{enumerate}
The determination of the background follows an approach different to
our earlier publications and will be described in more detail in Petry et al. (\cite{agnpaper}).
In order to maximise our exposure time at small
zenith angles the data were recorded in consecutive ON-source runs. 
OFF-source observations required for background determination were made
when Mkn 501 was not observable. Observations of 9 different ``empty-sky''
regions made before, during and after the Mkn 501 observing
season were available, forming a combined OFF-source dataset of 86.3 h at $5^\circ <
\theta < 25^\circ$. From these data the background was determined,
both for the Crab Nebula and for Mkn 501, as follows. 

From Monte Carlo studies we expect less than 1\% of source $\gamma$ events
in our camera to fulfill the condition $20^\circ <$ ALPHA $< 80^\circ$.
The number of events which pass all other cuts
and lie in this ALPHA region is therefore used to normalise the ALPHA-distribution 
of the OFF data to that of the ON data. 
Since the characteristics of the shower images are zenith angle ($\theta$) dependent, 
we adjust the $\theta$ distribution of the OFF data to that of the ON data,
by performing the normalization in 
$n$ separate $\theta$ bins.
The normalisation constants $\beta_1, ..., \beta_n$ are calculated using
\begin{quotation}
\begin{center}
\vspace{-0.3cm}
\begin{displaymath}
	\beta_i = \frac{M_{\mathrm{on},i}}{M_{\mathrm{off},i}}
\end{displaymath}
\end{center}
\end{quotation}
where $M_{\mathrm{on},i}$ is the number of ON-source events with
$20^\circ <$ ALPHA $< 80^\circ$ in $\theta$ bin $i$ and
$M_{\mathrm{off},i}$ is the corresponding number for the OFF data.
The width of the $\theta$ bins was a compromise between the accuracy
of a small bin width and sufficient events per bin for a low statistical error. 
For both Mkn 501 and the Crab Nebula $n$ = 3 approximately equidistant bins
between $\theta \approx 5^\circ$ and $\theta = 25^\circ$ were used.

The number of expected background events in the signal region of ALPHA $< 10^\circ$, $B$, is then obtained from
\begin{quotation}
\begin{center}
\vspace{-0.3cm}
\begin{displaymath}
	B = \sum_{i = 1}^{n} \beta_i N_{\mathrm{off},i}
\end{displaymath}
\end{center}
\end{quotation}
where $N_{\mathrm{off},i}$ is the number of OFF-source $\gamma$ candidates
(after all cuts including ALPHA $< 10^\circ$) in $\theta$ bin $i$.
By standard error propagation, the statistical error on $B$ is 
\begin{quotation}
\begin{center}
\vspace{-0.3cm}
\begin{displaymath}
	\sigma(B) = \sqrt{ \sum_{i = 1}^{n} \beta_i^2 N_{\mathrm{off},i} + (\beta_i^2 +\beta_i^3)N_{\mathrm{off},i}^2/M_{\mathrm{on},i} }
\end{displaymath}
\end{center}
\end{quotation}
We calculate the significance $S$ of the signal as the excess divided
by the statistical error on the excess:
\begin{quotation}
\begin{center}
\vspace{-0.3cm}
\begin{displaymath}
	S = \frac{N_{\mathrm{on}} - B}{\sqrt{N_{\mathrm{on}} + (\sigma(B))^2}}
\end{displaymath}
\end{center}
\end{quotation}
where $N_{\mathrm{on}}$ is the number of events in the ON-source dataset
after all cuts including ALPHA $< 10^\circ$.

This conservative approach takes into account both the dependency of the
image parameters on $\theta$ and the statistical error
in our knowledge of this dependency. 

\vspace{-0.1cm}
\section{Results}

\begin{figure}
\vspace{-0.5cm}
	 \epsfxsize=9.0cm
	 \epsffile{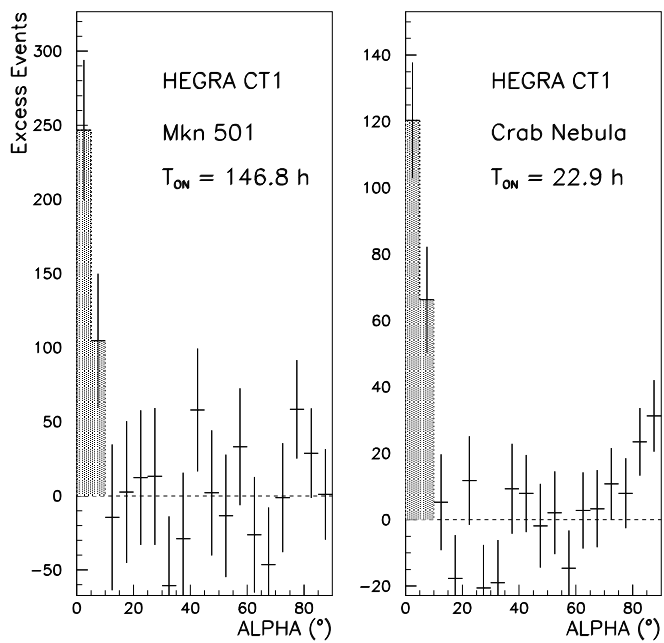}
         \vspace{-0.9cm}
	 \caption[]{\small
	 \label{alphaplot} The ON-source - OFF-source ALPHA
	 distributions. ALPHA is a measure of the deviation of the shower axis 
         from the source direction. For Mkn 501 there are 351 excess $\gamma$
	 candidates after all cuts (shaded, rate 2.4 h$^{-1}$ with a significance of
	 5.2 $\sigma$).  For the Crab Nebula there are 186 excess
	 $\gamma$ candidates (8.1 h$^{-1}$) with a significance of
	 7.6 $\sigma$.}
  \vspace{-0.8cm}
\end{figure}

\vspace{-0.1cm}
Table \ref{totsig} shows the event statistics. 
The number of expected events and the statistical significances are calculated according to the
procedure described in Section \ref{analysis}. We obtain an excess of 351 events after all cuts with a significance
of 5.2 $\sigma$.
For comparison, 23 h of observations of the Crab Nebula taken between October 1995 and February 1996 at $\theta$ 
$< 25^\circ$ were analysed. Using the same procedure as for Mkn 501, we find an excess of 186 events 
with a significance of 7.6 $\sigma$.
The ALPHA distributions for Mkn 501 and the Crab Nebula are shown in Figure \ref{alphaplot}.

\begin{table}
\begin{center}
\small{
  \begin{tabular}{|c|c|c|c|}
\hline
      &  & Crab & OFF \\
 \rb{target } 	& \rb{Mkn 501}	& Nebula & source \\
\hline
observation time (h) & 146.8	& 22.9	& 86.3	\\
\hline
events after filter & 359160	& 67045	& 209037	\\
\hline
events after all & 	& 	& 	\\
cuts except ALPHA & \rb{8083}	& \rb{1971}	& \rb{4633}	\\
\hline
{\bf events after all }& 	& 	& 	\\
{\bf cuts }	& \rb{1325}	& \rb{423}	& \rb{612}	\\
\hline
{\bf expected} background & 	&	&		\\
events\footnotemark[1] after all cuts  & \rb{974 $\pm$ 57}	& \rb{237 $\pm$	13}	&	\rb{-}	\\
\hline
excess & 351	& 186	&   -   \\
\hline
{\bf significance} & 	&  	&      \\
{\bf of excess} & \rb{5.2 $\sigma$}	& \rb{7.6 $\sigma$}	& \rb{-} \\
\hline
\end{tabular}
}
\caption{\label{totsig} Statistics of the Mkn 501, Crab Nebula and OFF source datasets. The number of expected events after all cuts is found from the
OFF source data ($B \pm \sigma(B)$ in Section \protect \ref{analysis}).}
\vspace{-0.5cm}
\end{center}
\end{table}
\footnotetext[1]{Note that the errors are different from the normal $\sqrt{N}$ since they  were obtained from the procedure described in
Section \ref{analysis}.}

\vspace{-0.2cm}
\section{Spectral Index and Flux calculations}
\label{mcstudy}

\vspace{-0.1cm}
The procedure to examine the spectral
behaviour of the source and subsequently determine the integral
photon flux follows largely the one described in Paper I.
Again, Monte Carlo (MC) data was produced using the code
described by Konopelko et al. (\cite{konopelko-b}).

      \begin{figure}
	 \vspace{0.1cm}
	 \epsfxsize=8.8cm
	 \epsffile{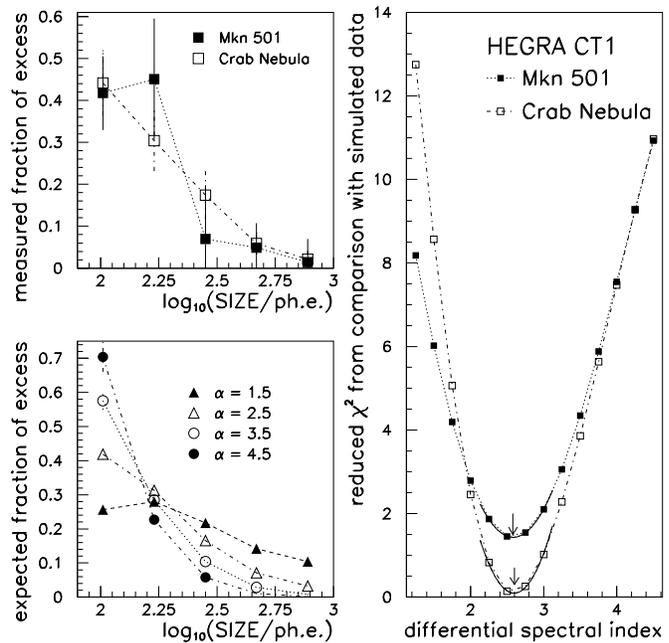}
         \vspace{-0.3cm}
         \caption[]{\small \label{specdist} {\bf a} (upper left) the measured differential SIZE distribution
	    for Mkn 501 and the Crab Nebula. The integrals of all distributions are normalized to 1.
	    The SIZE parameter is measured in units of photoelectrons (ph.e.). The errorbars are the combined statistical and
            systematic error.
	    {\bf b} (lower left) four examples of simulated SIZE spectra for different shapes
	    of the primary energy spectrum. $\alpha$ is the differential spectral index of the power law.
            {\bf c} (right) the reduced $\chi^2$ from the comparison of the measured with the simulated
            SIZE spectra as a function of $\alpha$.}
         \vspace{-0.4cm}
      \end{figure}

The effective collection area of CT1 for our given set of cuts was
calculated taking into account the dependence of the cut efficiencies
on primary energy and impact parameter. The main sources of systematic
errors in the flux calculation are the energy calibration, i.e. the
conversion from Cherenkov photons to ADC counts which our measurements
suggest is uncertain by $\pm$ 20 \%, and the determination of
the shape of the primary photon energy spectrum.

Spectral shape is estimated from analysis of the differential
distribution of the parameter SIZE which is defined as the total light
content of the shower image in the camera. Our MC simulations show
that here SIZE is in first
order proportional to the energy of the primary $\gamma$-ray, but fluctuates for
individual showers by up to 50 \% (2$\sigma$).

The measured distributions
for Mkn 501 and the Crab Nebula are shown in Figure \ref{specdist}a.
The signal in each SIZE bin was derived in the same way as the total signal
(see Section \ref{analysis}).
From MC simulations of $\gamma$-showers and the detector, the
shape of the SIZE distribution can be predicted for a given
night sky background noise, average zenith angle and shape of
the primary energy spectrum. Figure \ref{specdist}b
shows examples for power law spectra with differential spectral
index $\alpha$ = 1.5, 2.5, 3.5 and 4.5.

By varying $\alpha$ between 1.25 and 4.5 in steps of 0.25
and calculating the reduced $\chi^2$ of the comparison of the simulated
with the measured SIZE distribution, we find that the spectrum of the
Crab Nebula is compatible with an index 
\begin{quotation}
\begin{center}
\vspace{-0.3cm}
    $\alpha$(Crab Nebula) = $2.60 (\pm0.41)_{\mathrm{Stat}}$
\end{center}
\end{quotation}
The minimum reduced $\chi^2$ reached is 0.1 independent of whether
only the first three, four or all five points are used.
In order to test the robustness of the procedure we varied
the photon to ADC count conversion factor
by $\pm$ 20 \%. The most probable index changed by less than 0.1.

Assuming a spectral index of $2.6 \pm 0.4$, we determine the integral flux of
the Crab Nebula above 1.5 TeV from the 1995/96 CT1 data to be
\begin{quotation}
\begin{center}
    $7.7 (\pm1.0)_{\mathrm{Stat}} (+4.6-0.9)_{\mathrm{Syst}} \times 10^{-12}$ cm$^{-2}$s$^{-1}$
\end{center}
\end{quotation}
where the systematic error combines the uncertainties in the spectral index and
the energy calibration.

The same method applied to Mkn 501 results in a similar spectral index:
\begin{quotation}
\begin{center}
    $\alpha$(Mkn 501) = $2.58 (\pm0.51)_{\mathrm{Stat}}$
\end{center}
\end{quotation}
The minimum $\chi^2$ reached is 1.4.

Given this uncertainty in the spectral shape, we cannot make strong statements about
modifications to the power law spectrum such as cutoffs.
We find that the impact of a possible cutoff on our calculated
integral flux would be negligible if it was at energies $>$ 6 TeV.
We assume for the calculation
of the flux a spectral index of 2.6$\pm0.5$. Based on the total
Mkn 501 dataset at $\theta$ $< 25^\circ$ we thus find an integral
flux above 1.5 TeV of
\begin{quotation}
\begin{center}
 $2.3 (\pm0.4)_{\mathrm{Stat}} (+1.5-0.6)_{\mathrm{Syst}} \times 10^{-12}$ cm$^{-2}$s$^{-1}$
\end{center}
\end{quotation}

\vspace{-0.1cm}
\section{Variability}

\begin{figure}
         \vspace{-0.3cm}
	 \epsfxsize=9.0cm
	 \epsffile{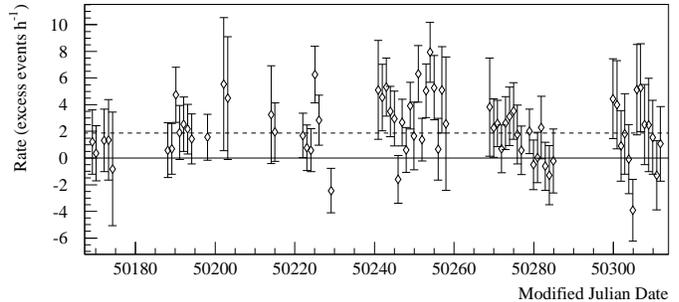}
         \vspace{-0.7cm}
         \caption[]{\small \label{dailyrate}
	Daily excess $\gamma$-ray event rates from Mkn 501. A constant fit (dashed line) to 70 nights of CT1 data 
        (the 70 nights on which Mkn 501 was observed at $\theta$ $< 25^\circ$ for at least 20 minutes) 
        leads to a rate of 1.88 $\pm$ 0.26 h$^{-1}$, consistent within our errors with the mean rate obtained 
        from the total $\theta$ $< 25^\circ$ dataset of 2.39 $\pm$ 0.46 h$^{-1}$. The peak rate of 7.9 $\pm$ 2.2  h$^{-1}$, recorded during
	a 3.5 h observation centred on MJD 50253.97, lies 2.5 standard deviations from the mean.}
        \vspace{-0.2cm}
  \end{figure}

With this Mkn 501 dataset collected over an exceptionally long period (approx. 6 months) during which
no significant change was made to the telescope and data processing stream, an investigation of the stability of
our signal over time is of interest. In fact, flux variability is a common
characteristic of EGRET blazar measurements (von Montigny et al. \cite{montigny}). The Whipple group finds the VHE flux of Mkn 501's 
sister object Mkn 421 to be highly variable on timescales of months to days
(Macomb et al. \cite{macomb}) even observing a flare event in which the flux rose to $>$ 50 times its quiescent level (Gaidos et al. \cite{gaidos}). During the 1995 Whipple Mkn 501 observations 
the flux from this source appeared approximately constant (Quinn et al. \cite{quinn}) except on one day
(of 58) when the measured rate lay 5 standard deviations above the mean. 
Figure \ref{dailyrate} shows the rate of excess $\gamma$-ray candidate events recorded by CT1 on nights when Mkn 501 was observed for more than 
20 minutes at $\theta$ $< 25^\circ$. Above our threshold of 1.5 TeV we have seen no significant ``burst behaviour''.

\vspace{-0.2cm}
\section{Conclusion}

\vspace{-0.15cm}
We have detected $\gamma$-radiation above 1.5\ TeV from Mkn 501
using the first imaging air Cherenkov telescope of the HEGRA Collaboration. 
An integral flux above 1.5 TeV of approx. $2.3\times 10^{-12}$ cm$^{-2}$s$^{-1}$
was found from the March - August 1996 data.  Mkn 501
shows a spectrum compatible with a differential spectral index of 2.6$\pm 0.5$, where the
large error is mainly statistical.

Our near contemporary Crab Nebula data show an excess rate 3.4 times larger than that for Mkn 501
and a spectral index of 2.6$\pm0.4$, from which a flux above
1.5 TeV of approx. $7.7 \times 10^{-12}$ cm$^{-2}$s$^{-1}$ is calculated.
This is in very good agreement with our earlier measurement using CT2 at
a threshold of 1 TeV
(Petry et al. \cite{421paper}).
Within the statistical errors,
the integral flux above 1.5 TeV from Mkn 501 is less than that from the Crab Nebula by a factor of 2.9$\pm 0.7$.

In 70 nights of data no significant deviation of the Mkn 501 event rate from a mean of 2.4 events h$^{-1}$ was observed. 

\vspace{-0.2cm}
\section*{Acknowledgements}

\vspace{-0.05cm}
The HEGRA Collaboration thanks the Instituto de Astrofisica de Canarias for 
use of the HEGRA site at the Roque de los Muchachos and its facilities. This work was supported
by the BMBF, the DFG and the CICYT.

\end{document}